Xie and Wang *Financial Innovation* (2018) 4:30
https://doi.org/10.1186/s40854-018-0110-4

Financial Innovation

# RESEARCH

**Open Access**

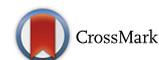

# Timing the market: the economic value of price extremes

Haibin Xie[1]  and Shouyang Wang[2*]

* Correspondence:
sywang@amss.ac.cn
[2]Academy of Mathematics and
Systems Science, Chinese Academy
of Sciences, Beijing 100190, China
Full list of author information is
available at the end of the article

## Abstract

By decomposing asset returns into potential maximum gain (PMG) and potential maximum loss (PML) with price extremes, this study empirically investigated the relationships between PMG and PML. We found significant asymmetry between PMG and PML. PML significantly contributed to forecasting PMG but not vice versa. We further explored the power of this asymmetry for predicting asset returns and found it could significantly improve asset return predictability in both in-sample and out-of-sample forecasting. Investors who incorporate this asymmetry into their investment decisions can get substantial utility gains. This asymmetry remains significant even when controlling for macroeconomic variables, technical indicators, market sentiment, and skewness. Moreover, this asymmetry was found to be quite general across different countries.

## Introduction

It is well known that price extremes contain valuable information for estimating and forecasting the volatility of financial assets. Parkinson (1980), Beckers (1983), Garman and Klass (1980), Wiggins (1991), Rogers and Satchell (1991), Kunitomo (1992), and Yang and Zhang (2000), among others, demonstrated the superiority of using price range (defined as the difference between high and low extreme prices) as a volatility estimator as compared with standard methods. Sassan et al. (2002) show that a range-based volatility estimator is not only highly efficient but also approximately Gaussian and robust to microstructure noise. Chou (2005) proposed the conditional autoregressive range model (CARR) and found that it provided sharper volatility estimates compared to a standard GARCH model. Brandt and Jones (2006) proposed a range-based EGARCH model and found substantial forecastability of volatility. Martens and Dijk (2007) showed that realized range is a more efficient estimator of volatility than realized volatility.

It remains unknown whether price extremes contribute to forecasting asset returns. Despite a great deal of research on range-based volatility, few studies, to our knowledge, have related asset returns to price extremes. In an intriguing study, George and Hwang (2004) noted that traders appear to use the 52-week high as a reference point against which they evaluate the potential impact of news. Hence, nearness to the 52-week high is positively associated with expected returns in the cross section. Further, Li and Yu (2012) suggested traders may use the historical high as another anchor against which they evaluate information. They also showed that the 52-week high and the historical high contain information about future market returns that is not captured by traditional macroeconomic variables. George and Hwang (2004) and Li and Yu (2012) both hinted that price extremes might have additional information in explaining asset returns.

This study attempted to relate asset returns to price extremes through intuitive decomposition. The next section will show that asset returns can be decomposed into potential maximum gain (PMG) and potential maximum loss (PML). The idea of decomposing asset returns into PMG and PML was motivated by both psychological and empirical findings. Mounting evidence shows that investors have different and asymmetric reactions to gains/good news and losses/bad news. Kahneman and Tversky (1979), for example, found that "the value function is normally concave for gains, commonly convex for losses, and is generally steeper for losses than for gains." This finding indicates that traders' reactions to good news should be different from their reactions to bad news. Veronesi (1999) showed that with correct beliefs, rational investors may overreact to bad news in good times and underreact to good news in bad times when the market shifts between two unobservable states. Andersen et al. (2003) used a new dataset of exchange rate quotations, macroeconomic expectations, and macroeconomic realizations to explore real-time price discovery in a foreign exchange. They found that exchange markets react to news in an asymmetric way, and bad news has a greater impact than good news. Nguyen and Claus (2013) explored heterogenous consumers' reactions to a range of financial and economic news and found asymmetry in the responses to news, where consumers reacted to bad news but not good news. Thus, we expect that decomposing asset returns into PMG and PML could provide a new and different profile of the dynamics of asset returns.

We found that both PMG and PML displayed very interesting time series properties. First, there is high persistence in both PMG and PML. It is well documented that asset returns, especially for monthly and quarterly observations, display no or low persistence. However, we found significant persistence in PMG and PML, and PML is more persistent than PMG. This finding seems to be consistent with Hong and Stein (1999), who claimed that information diffuses gradually, and with Hong, Lim, and Stein (2000), who reported that bad news travels slowly. Second, we found a significant asymmetry between PML and PMG. The Granger causality test (Granger, 1969) showed that PML caused PMG but not vice versa. Further empirical evidence showed that PML positively predicted PMG, which means that a larger potential loss implies future large potential gains. This finding suggests that investors overreact to bad news.

The asymmetry found in this study is valuable for timing the market. We found that such asymmetry can be used to improve the predictability of stock returns in both in-sample and out-of-sample forecasting. Financial economists have sought to identify variables that forecast aggregate stock market returns. Welch and Goyal (2008) found that a long list of predictors from the literature was unable to deliver consistently superior out-of-sample forecasts relative to a simple forecast based on historical average. Thus, it would be of great interest to investigate the forecasting power of this asymmetry. We found a sharp increase in return predictability in both in-sample and out-of-sample forecasting once asymmetry was considered. For monthly data, the in-sample R-square ranged from 1.57% to 1.89%, and the out-of-sample R-square ranged from 0.098% to 1.46%. Similar results were obtained for quarterly data. The R-square ranged from 2.60% to 12.54% for in-sample fitting and from 0.97% to 3.87% for out-of-sample forecasting.

We also investigated the economic value of this asymmetry and found that it can provide substantial utility gains. Suppose a mean-variance investor with a risk aversion of 3 invests over the market portfolio and the Treasury bill. Using asymmetry to predict the market return, the investor can obtain 0.378%–2.41% (for monthly horizon) and 2.95%–4.85% (for quarterly horizon) more annualized certainty equivalent return (CER) relative to the strategy using the historical return average as the expected market return estimate. The Sharpe ratio provides further evidence that this asymmetry generates economic value.

The asymmetry between PMG and PML cannot be explained by business-cycle-related variables, technical indicators, market sentiment, and skewness. Rapach, Strauss, and Zhou (2010); Henkel, Martin, and Nardari (2011); and Dangl and Halling (2012), among others, found significant asymmetry in return forecasting. Macroeconomic variables usually show significant out-of-sample forecasting in bad times but insignificant or weaker forecasting in good times. However, we found that asymmetry cannot be attributed to business cycle. Asymmetry remains significant, even when controlling for the commonly used macroeconomic variables. Baker and Wurgler (2006) found that market sentiment has strong forecasting power for a large number of cross-sectional stock returns. Baker, Wurgler, and Yuan (2012) provided further international evidence for the cross-section forecasting power of investor sentiment. Huang et al. (2016) presented two technical indicators that significantly predict stock returns: the mean reversion indicator and the good state indicator. Huang et al. (2015) provided evidence of market sentiment predicting aggregate stock returns. However, we found that asymmetry cannot be explained by either market sentiment or technical indicators. We also investigated whether asymmetry can be explained by skewness in stock returns and found it cannot.

This study is related to George and Hwang (2004) and Li and Yu (2012). However, we differ from them in at least three aspects. First, both 52-week high and historical high were constructed from closing price. In this paper, the *high price extreme* refers to the highest trading price over a specified time interval. Second, both 52-week high and historical high served mainly as proxies for news levels. For example, if nearness to the 52-week high is high, it is more likely that the firm experienced good news in the recent past. That is, 52-week high and historical high are used to measure the certainty of the news. In this study, the highest trading price was used as a proxy for news uncertainty. The higher the highest trading price, the more uncertainty in the price changes. Third, George and Hwang (2004) and Li and Yu (2012) focused on the effect of investors' psychology on asset pricing. This study focused instead on the time series properties of asset returns.

The main contributions of this paper are summarized as follows. First, we present new evidence confirming the economic value of price extremes in forecasting asset returns. Second, we document a new asymmetry in asset returns; PML has a larger impact on PMG than PMG has on PML. This asymmetry could be used to improve return predictability.

The rest of the paper is organized as follows. Section 2 describes the empirical methodology. Section 3 provides empirical results showing significant asymmetry between PMG and PML. Section 4 shows the power of asymmetry for predicting asset returns along with its economic value in investment. Section 5 presents potential explanations for asymmetry. Section 6 presents global evidence for asymmetry across the main stock indices. Section 7 concludes the paper.

## Econometric Methodology

### Return Decomposition

Traditionally, the literature on stock returns has been exclusively based on closing price:

$$r_t = \ln(C_t) - \ln(C_{t-1}), \qquad (1)$$

where $C_t$ is the closing price at time $t$, and $r_t$ is the logarithmic return over a holding period from $t$-1 to $t$.

A problem with Equation (1) is that it ignores price movements from time $t$-1 to $t$, which means there is missing information. To alleviate this problem, we propose decomposing the stock returns with the high price extreme:

$$\begin{aligned} r_t &= \ln(C_t) - \ln(C_{t-1}) \\ &= [\ln(O_t) - \ln(C_{t-1})] + [\ln(H_t) - \ln(O_t)] - [\ln(H_t) - \ln(C_t)] \\ &= OVR_t + PMG_t - PML_t, \end{aligned} \qquad (2)$$

where $O_t$ and $H_t$ are, respectively, the opening price and the high price over [t-1, t]. This shows that stock returns over [t-1, t] comprise three components:

- *Overnight returns* (OVR$_t$). $OVR_t = ln(O_t) - ln(C_{t-1})$. The overnight return gauges the return due to overnight information.
- *Potential maximum gain* (PMG$_t$). $PMG_t = ln(H_t) - ln(O_t)$. The potential maximum gain measures the possible maximum profit from the opening price to the high price extreme.
- *Potential maximum loss* (PML$_t$). $PML_t = ln(H_t) - ln(C_t)$. The potential maximum loss measures the possible maximum loss from the high price extreme to the closing price.

Equation (2) indicates not only the returns but also the equity risk. In this paper, we call Equation (2) the *return decomposition*. PMG and PML measure the uncertainty of price changes or the equity risk. From Equation (2), it can also be seen that PMG and PML can be used as proxies for good news and bad news. Therefore, the time series dynamics of PMG and PML describe how good news and bad news are incorporated into equity prices or the price-discovery process.

The return decomposition technique was mainly based on Kahneman and Tversky (1979), George and Hwang (2004), and Li and Yu (2012). Kahneman and Tversky (1979) showed that investors behave differently when facing possible gains and losses. George and Hwang (2004) and Li and Yu (2012) found that investors use high prices as anchors. Therefore, we conjecture that high price extremes can help us to better understand the dynamics of asset returns.

For data observations of low frequency, overnight returns contribute very little to variations in asset returns and thus can be neglected. In the next section, asset returns, unless specified otherwise, refer to returns with the overnight returns removed.

### Dynamics of Asset Returns

For time series data, the most commonly used econometric tool is covariance analysis. The covariance between asset returns $r_t$ and $r_{t-i}$ can be presented as follows:

$$\begin{aligned} Cov(r_t, r_{t-i}) &= Cov(PMG_t - PML_t, PMG_{t-i} - PML_{t-i}) \\ &= [Cov(PMG_t, PMG_{t-i}) + Cov(PML_t, PML_{t-i})] \\ &\quad - [Cov(PMG_t, PML_{t-i}) + Cov(PML_t, PMG_{t-i})] \end{aligned}$$

This equation shows that the covariance in asset returns is determined by two parts: the autocovariances in $PMG_t$ and $PML_t$ ($Cov(PMG_t, PMG_{t-i})$, $Cov(PML_t, PML_{t-i})$), and the cross covariances between $PMG_t$ and $PML_t$ ($Cov(PMG_t, PML_{t-i})$, $Cov(PML_t, PMG_{t-i})$). Each part has an economic sense. The autocovariances in $PMG_t$ and $PML_t$ measure, to some extent, the persistence of good news and bad news, respectively. The larger the autocovariance, the more slowly news travels. The cross covariances between $PMG_t$ and $PML_t$ measure the interactions between $PMG_{t-i}$ ($PML_{t-i}$) and $PML_t$ ($PMG_t$). Therefore, the return decomposition shows that the time series dynamics of asset returns have very complicated and subtle intrinsic structures.

We modeled the autocovariance in PMG and PML with an ARMA(l, m)-GARCH(p, q) model:

$$\begin{aligned} S_t &= \mu + \sum_{i=1}^{l} \phi_i S_{t-i} + \sum_{j=1}^{m} \theta_j \mu_{t-j} + \mu_t \\ \mu_t &= \sigma_t \upsilon_t \\ \sigma_t^2 &= \omega + \sum_{i=1}^{p} \alpha_i \sigma_{t-i}^2 + \sum_{j=1}^{q} \beta_j \mu_{t-j}^2 \\ \upsilon_t &\sim N(0,1), \text{i.i.d}, \end{aligned} \quad (3)$$

where $S_t = PMG_t, PML_t$. GARCH(p, q) was used because financial markets are notoriously well known for their heteroscedasticity.

The Granger causality test (Granger, 1969) test was applied to the cross covariance. The Granger causality test is used to determine whether one time series is useful for forecasting another. A time series $X$ is said to Granger-cause $Y$ if it can be shown that those $X$ values provide statistically significant information about future values of $Y$. The test for causality in the Granger sense is based on the following equations:

$$Y_t = \alpha_0 + \sum_{j=1}^{m} \alpha_j Y_{t-j} + u_t \quad (4)$$

$$Y_t = \beta_0 + \sum_{j=1}^{m} \beta_j Y_{t-j} + \sum_{i=1}^{m} \gamma_i X_{t-i} + v_t \quad (5)$$

where $u_t$ and $v_t$ are independent, series-uncorrelated random variables with zero means and finite variances. Whether $X$ Granger-causes $Y$ is based on a test of the null hypothesis that $\gamma_1 = \gamma_2 = \ldots = \gamma_n = 0$. Rejection of the null hypothesis means $X$ causes $Y$ in the Granger sense.

Analyzing the cross covariance between $PMG_t$ and $PML_t$ is of greater interest. First, cross covariance between $PMG_t$ and $PML_t$ can be used to describe how investors form their expectations on future gains (losses) conditional on historical losses (gains). Thus, it is related to the literature on investors' asymmetric reactions to gains and losses. Second, it is highly related to return predictability. Predicting stock returns $r_t$ conditional on the historical information set can be presented as

$$E(r_t|\Omega_{t-1}) = E[(PMG_t - PML_t)|\Omega_{t-1}], \quad (6)$$

where $\Omega_t = \{r_t, r_{t-1}, ....\}$ Equation (6) shows that, unless investors predict $PMG_t$ and $PML_t$ using the same information and the same model, modeling stock returns as a unit may produce misleading results. For example, if PML Granger-causes PMG and not vice versa, then modeling stock returns as a unit is not equivalent to modeling PMG and PML:

$$E(r_t|\Omega_{t-1}) = E[(PMG_t - PML_t)|\Omega_{t-1}]$$
$$\neq E(PMG_t|\Omega^d_{t-1}) - E(PML_t|\Omega^d_{t-1}),$$
$$\text{where } \Omega^d_{t-1} = \{(PMG_t, PML_t), (PMG_{t-1}, PML_{t-1}), ....\}$$

To further quantify the interaction between PMG and PML, we performed the following regression:

$$\chi_t = c + \eta_i \psi_{t-i} + \varsigma_t, \quad (7)$$

where $\chi_t$ and $\psi_{t-i}$ are filtered $PMG_t$ or $PML_t$. The filtered $\chi_t$ and $\psi_{t-i}$ were obtained by first removing the autocorrelations in PMG and PML and then standardizing the residuals. The coefficient $\eta_i$ directly measures the impact of unit $\psi_{t-i}$ on $\chi_t$.

## Empirical Results

### Data

We collected the monthly index data of the Standard and Poor's 500 (S&P 500) for the sample period January 1950 to December 2015 with 792 observations. The data set was downloaded from the finance subdirectory of the website Yahoo.com.[1] For each month, four pieces of price information—opening, closing, high, and low—are reported. Since the website does not provide quarterly index data, we constructed quarterly index data from monthly observations. The construction is presented as follows:

$$L^q_t = Min_t\{L^m_{3(t-1)+1}, L^m_{3(t-1)+2}, L^m_{3t}\}, H^q_t = Max_t\{H^m_{3(t-1)+1}, H^m_{3(t-1)+2}, H^m_{3t}\},$$
$$O^q_t = O^m_{3(t-1)+1}, C^q_t = C^m_{3t}, t = 1, 2, 3, ...$$

The labels $q$ and $m$ represent, respectively, the quarterly and monthly observations. For quarterly index data, there were 264 observations. From the collected and constructed data, the stock returns, potential maximum gains (PMG), and potential maximum losses (PML) were calculated by their definitions.

Table 1 presents the summary statistics on the stock returns PMG and PML. The kurtosis coefficients of the stock returns PMG and PML on either monthly or quarterly observations are larger than 3, indicating a strong deviation from the normal distribution. It is interesting to observe the difference in the values of the *ACF*s and the Ljung-Box *Q* statistics for stock returns PMG and PML. The *Q* statistics for stock returns on either monthly or quarterly observations are small and statistically insignificant at the level of 10%, indicating no significant persistence in stock returns. Meanwhile, the *Q* statistics for PMG and PML are statistically significant at the level of 10%, indicating evidence of persistence in PMG and PML. The persistence in PMG and PML is consistent with Hong and Stein (1999), who suggested that information diffuses gradually. Consistent with Hong, Lim, and Stein (2000), who reported that bad news travels slowly, the *Q* statistics also show more persistence in PML than in PMG.

Table 2 presents the correlation statistics among stock returns ($r_t$), PMG, PML and stock return with overnight return included ($r°_t$). The high correlations between $r_t$ and

Table 1 Summary Statistics on Stock Returns, Potential Maximum Gains, Potential Maximum Losses

| | Panel A. Monthly Index Data | | | Panel B. Quarterly Index Data | | |
|---|---|---|---|---|---|---|
| | $r_t$ | $PMG_t$ | $PML_t$ | $r_t$ | $PMG_t$ | $PML_t$ |
| Mean | 6.051E-03 | 0.033 | 0.027 | 0.018 | 0.063 | 0.045 |
| Std.Dev | 0.042 | 0.025 | 0.029 | 0.078 | 0.045 | 0.051 |
| Maxi | 0.151 | 0.178 | 0.267 | 0.195 | 0.238 | 0.313 |
| Mini | -0.245 | 0.000 | 0.000 | -0.303 | 0.000 | 0.000 |
| Skew | -0.655 | 1.413 | 2.373 | -0.949 | 1.034 | 2.268 |
| Kurt | 5.435 | 6.358 | 12.771 | 4.920 | 4.00 | 9.143 |
| J-B stat | 251.9 | 635.7 | 3893.8 | 79.9 | 58.0 | 641.4 |
| Prob | 0.000 | 0.000 | 0.000 | 0.000 | 0.000 | 0.000 |
| | | Auto-Correlation Function (lag) | | | | |
| ACF(1) | 0.046 | 0.083 | 0.279 | 0.085 | 0.185 | 0.220 |
| ACF(3) | 0.043 | 0.176 | 0.169 | -0.042 | -0.064 | 0.059 |
| ACF(6) | -0.058 | 0.060 | 0.074 | -0.033 | -0.068 | 0.029 |
| ACF(9) | -0.021 | 0.062 | 0.091 | -0.004 | 0.010 | -0.063 |
| ACF(12) | 0.050 | 0.081 | 0.129 | 0.007 | -0.075 | -0.008 |
| Q(12) | 16.41 | 97.56*** | 203.81*** | 9.57 | 17.84* | 24.39** |
| Obs | 263 | 264 | 264 | 792 | 792 | 792 |

Note. J-B stat means the *Jarque-Bera* statistics. Q(12) represents the *Ljung-Box Q* statistics.***, **, * means respectively statistical significance at the level of 1%, 5% and 10%

$r°_t$ (0.997 for monthly observations and 0.999 for quarterly observations) indicate that $r°_t$ can be perfectly approximated by $r_t$. Regressing $r°_t$ on $r_t$, we found that $r_t$ can almost fully explain the variation of $r°_t$. The *R*-squares for monthly observations and quarterly observations were 99.4% and 99.8%, respectively, which means overnight returns contribute very little to variations in stock returns and thus can be omitted in our empirical analysis. The correlation between PMG and PML, instead of being uncorrelated, is reported to be significantly positive.

### Autocorrelations in PMG and PML

The summary statistics in Table 1 show that the distributions of PMG and PML are severely skewed and far from normal. In this study, we alleviated skewness by using squared root transformation on both PMG and PML. Possible heteroscedasticity was

Table 2 Correlation Analysis on Stock Returns ($r_t$), Stock Returns with Overnight Returns being included ($r_t^o$), $PMG_t$ and $PML_t$

| | Panel A. Monthly Index Data | | | | Panel B. Quarterly Index Data | | | |
|---|---|---|---|---|---|---|---|---|
| | $r_t^o$ | $r_t$ | $PMG_t$ | $PML_t$ | $r_t^o$ | $r_t$ | $PMG_t$ | $PML_t$ |
| $r_t^o$ | 1.000 | — | — | — | 1.000 | — | — | — |
| $r_t$ | 0.997*** | 1.000 | — | — | 0.999*** | 1.000 | — | — |
| $PMG_t$ | 0.732*** | 0.737*** | 1.000 | — | 0.772*** | 0.775*** | 1.000 | — |
| $PML_t$ | 0.802*** | 0.802*** | 0.187*** | 1.000 | 0.831*** | 0.830*** | 0.290*** | 1.000 |

Note.***, **, * means respectively statistical significance at the level of 1%, 5% and 10%. We regress $r_t^o$ on $r_t$, and the results are presented as follows.For monthly stock returns,
$r_t^o = 4.16E - 04 + 1.003 r_t + \varepsilon_t R^2 = 0.994$
For quarterly stock returns,
$r_t^o = 4.52E - 04 + 0.999 r_t + \varepsilon_t R^2 = 0.998$

also taken into consideration. We used Equation (3) to describe the dynamics of PMG and PML. Different ARMA(l, m)-GARCH(p, q) (l=1, 2; m=1, 2; p=1, q=1) models were used, and the final models were determined by the Akaike Information Criterion (AIC).

The modeling results are presented in Table 3. The results show heteroscedasticity in monthly observations but not in quarterly observations. It is interesting to note the differences in the R-square values. We found, for both quarterly and monthly observations, that PML was more predictable than PMG. The predictability of PML is almost twice that of PMG.

Table 4 presents the summary statistics on filtered PMG and PML. The results show that the kurtosis and J-B statistics decreased significantly after filtration compared to the results in Table 1. For PMG, the J-B statistics indicate that the null hypothesis of normal distribution cannot be rejected. The values of ACFs and Ljung-Box Q statistics for PMG and PML are small and statistically insignificant, indicating that the autocorrelations have been well filtered.

### Cross Correlation Between PMG and PML

Granger causality tests were employed to investigate the cross correlation between PMG and PML. Since Granger causality tests are sensitive to lags, different lags are used for robustness. We performed Granger causality tests on both unfiltered and filtered observations.

Table 5 reports the test results. For the unfiltered data, the null hypothesis that PML does not Granger-cause PMG is consistently rejected at the significance level of 5% for both monthly and quarterly data observations. The results for the null hypothesis that PMG does not Granger-cause PML are mixed. For monthly data, the null hypothesis is rejected when lag = 2; otherwise the null hypothesis cannot be rejected at the significance level of 5%. For quarterly data, the null hypothesis is rejected when lag = 2, 4. The results for filtered data are similar to the unfiltered ones, except that the null hypothesis that PMG does not Granger-cause PML cannot be rejected, and the null hypothesis that PML does not Granger-cause PMG is rejected.

This finding is interesting as it indicates an asymmetry between PMG and PML. For monthly data, the historical PML helps to predict PMG but not vice versa; for quarterly

Table 3 Autocorrelation Analysis on PMG and PML

| Filtered | Panel A. Monthly Index Data | | Panel B. Quarterly Index Data | |
| --- | --- | --- | --- | --- |
| | $Sqrt(PMG_t)$ | $Sqrt(PML_t)$_ | $Sqrt(PMG_t)$ | $Sqrt(PML_t)$ |
| μ | 0.164*** | 0.143*** | 0.020*** | 0.183*** |
| AR(1) | 0.891*** | 0.953*** | 0.152** | 0.666*** |
| MA(1) | -0.895*** | -0.841*** | | -0.460*** |
| AR(2) | | | | |
| MA(2) | 0.110*** | | | |
| ω | 0.352E-03 | 0.420E-03 | | |
| ARCH(1) | 0.035* | 0.060** | | |
| GARCH(1) | 0.895*** | 0.868*** | | |
| R-squared(%) | 5.45 | 12.11 | 2.32 | 7.05 |

Note: ***, **, * means respectively statistical significance at the level of 1%, 5% and 10%. Due to their high skewness and kurtosis, we perform squared root transform on both good extreme returns and bad extreme returns before filtration

Table 4 Summary Statistics on Filtered PMG and PML

| Filtered | Panel A. Monthly Index Data | | Panel B. Quarterly Index Data | |
|---|---|---|---|---|
| | $PMG_t$ | $PML_t$ | $PMG_t$ | $PML_t$ |
| Mean | 0.000 | 0.000 | 0.000 | 0.000 |
| Std.dev | 1.000 | 1.000 | 1.000 | 1.000 |
| Maxi | 3.637 | 4.915 | 2.515 | 3.329 |
| Mini | -2.748 | -2.343 | -2.383 | -1.933 |
| Skew | -0.006 | 0.538 | 0.098 | 0.740 |
| Kurt | 3.212 | 3.312 | 2.564 | 3.363 |
| J-B stat | 1.486 | 41.4 | 2.505 | 26.346 |
| Prob | 0.476 | 0.000 | 0.286 | 0.000 |
| Auto-Correlation Function (lag) | | | | |
| ACF(1) | 0.003 | 0.039 | 0.010 | 0.008 |
| ACF(3) | 0.039 | -0.013 | -0.021 | -0.057 |
| ACF(6) | -0.026 | -0.056 | -0.064 | -0.038 |
| ACF(9) | -0.008 | -0.020 | 0.017 | -0.055 |
| ACF(12) | 0.035 | 0.053 | -0.029 | 0.047 |
| Q(12) | 12.29 | 12.042 | 1.092 | 9.036 |
| Obs | 791 | 791 | 263 | 263 |

Note. To make sure zero sample mean and unit sample standard deviation, the stock returns, good extreme returns and bad extreme returns are first filtered, and then standardized using the following formula
$Z_t = [X_t - M(X_t)]/S(X_t)$
where $M(X_t)$ is the sample mean of $X_t$, $S(X_t)$ is the sample standard deviation of $X_t$

data, the historical PML contributes more to forecasting PMG than PMG does to PML. This asymmetry suggests that modeling stock returns as a unit might fail to discover the true time series dynamics of stock returns.

The quantitative interactions between PMG and PML were also calculated using Equation (7). We only report the results when $i = 1$. For filtered monthly observations,

$$PMG^F_t = 0.001 + 0.136^{***}PML^F_{t-1} + \varsigma_t$$
$$PML^F_t = 1.15E\text{-}04 - 4.484E\text{-}03 PMG^F_{t-1} + \varsigma_t.$$

For quarterly observations,

Table 5 Granger Causality Tests on PMG and PML. Decomposition with High Price Extremes

| | Panel A. Monthly Data Observations | | | Panel B. Quarterly Data Observations | | |
|---|---|---|---|---|---|---|
| Lags | 2 | 4 | 6 | 2 | 4 | 6 |
| PMG /↛ PML | 0.009 | 0.712 | 0.878 | 0.036 | 0.049 | 0.107 |
| PML /↛ PMG | 0.000 | 0.000 | 0.000 | 0.000 | 0.000 | 0.000 |
| $PMG^F$ /↛ $PML^F$ | 0.493 | 0.794 | 0.869 | 0.014 | 0.007 | 0.029 |
| $PML^F$ /↛ $PMG^F$ | 0.000 | 0.000 | 0.000 | 0.000 | 0.000 | 0.000 |

Note. $X$ /↛ $Y$ means the null hypothesis that $X$ does not Granger-causes $Y$. $PMG^F$ and $PML^F$ mean filtered PMG and PML respectively. This table reports the p-values of the F-statistics. When performing Granger causality test, we set $m = n$ in Equation (5) for the sake of being convenient. Different lags are used for being robust since the Granger causality tests are sensitive to the lag selection. Panels A and B report respectively the results for monthly and quarterly data observations. Decomposing stock returns with high price extremes are presented as
$r_t = [log(H_t) - log(O_t)] - [log(H_t) - log(C_t)]$
$= PMG_t - PML_t$

**Table 6** Granger Causality Tests on PMG and PML. Decomposition with Low Price Extremes

| Lags | Panel A. Monthly Data Observations | | | Panel B. Quarterly Data Observations | | |
| --- | --- | --- | --- | --- | --- | --- |
| | 2 | 4 | 6 | 2 | 4 | 6 |
| PMG /↛ PML | 0.390 | 0.851 | 0.622 | 0.594 | 0.566 | 0.436 |
| PML /↛ PMG | 0.000 | 0.000 | 0.000 | 0.000 | 0.000 | 0.000 |
| $PMG^F$ /↛ $PML^F$ | 0.967 | 0.952 | 0.908 | 0.852 | 0.792 | 0.390 |
| $PML^F$ /↛ $PMG^F$ | 0.000 | 0.000 | 0.000 | 0.002 | 0.004 | 0.014 |

Note. X /↛ Y means the null hypothesis that X does not Granger-causes Y. $PMG^F$ and $PML^F$ mean filtered PMG and PML respectively. This table reports the p-values of the F-statistics. When performing Granger causality test, we set m = n in Equation (5) for the sake of being convenient. Different lags are used for being robust since the Granger causality tests are sensitive to the lag selection. Panels A and B report respectively the results for monthly and quarterly data observations. Decomposing stock returns with high price extremes are presented as
$r_t = [\log(C_t) - \log(L_t)] - [\log(L_t) - \log(O_t)]$
$= PMG_t - PML_t$

$$PMG^F_t = -3.843E\text{-}03 + 0.330^{***}PML^F_{t-1} + \varsigma_t$$
$$PML^F_t = -6.171E\text{-}03 - 0.168^{***}PMG^F_{t-1} + \varsigma_t.$$

F represents filtered data observations.

The regression results confirmed the asymmetric impact between PMG and PML. PML had larger impact on PMG. For monthly data, the slope coefficients showed that one unit shock of PML increased PMG by 13.6% while PMG had almost no impact on PML. For quarterly data, the slope coefficients showed that one unit shock of PML increased PMG by 33.0% while one unit shock of PMG decreased PML by 16.8%.

The regression results suggest that stock markets overreact to bad news and underreact to good news, especially for quarterly data. The reasons are presented as follows. The positive impact of PML on PMG (the slope coefficients are reported to be positive) means that large potential losses imply future high potential gains—that is, the stock market overreacts to bad news. The negative impact of PMG on PML (the slope coefficients are reported to be negative) means that large potential gains hint at low potential losses—that is, the stock market underreacts to good news.

It might be argued that our Granger causality tests results are attributable to decomposing stock returns with high price extremes. The reasoning might be that $PML_{t-1}$ is followed by $PMG_t$ while $PML_t$, though following $PMG_{t-1}$, is interrupted by $PML_{t-1}$ and $PMG_t$. If this reasoning holds, then decomposing stock returns with low price extremes tends to result in Granger causality from PMG to PML. Decomposing stock returns with low price extremes is presented as follows:

$$r_t = [\ln(C_t) - \ln(L_t)] - [\ln(O_t) - \ln(L_t)] \\ = PMG_t - PML_t, \qquad (8)$$

where $L_t$ is the low price. To be robust, we also performed Granger causality tests between PMG and PML using Equation (8); the results are presented in Table 6. The results reported in Table 6 consistently show significant evidence of Granger causality from PML to PMG but not vice versa.

### Economic Value of Price Extremes

An interesting question to consider is, "Does it make any difference, especially economic difference, if asymmetry is taken into consideration?"

**Table 7** Vector AutoRegressive Model Estimation

| Panel A. Monthly Data Observations | | | | | | |
|---|---|---|---|---|---|---|
| | Over 1950.01-1985.12 | | Over 1986.01-2015.12 | | Over 1950.01-2015.12 | |
| | $PMG_{t+1}$ | $PML_{t+1}$ | $PMG_{t+1}$ | $PML_{t+1}$ | $PMG_{t+1}$ | $PML_{t+1}$ |
| $PMG_t$ | -0.028 [-0.573] | 0.065 [1.245] | 0.067 [1.213] | 0.032 [0.432] | -0.005 [-0.146] | 0.027 [0.615] |
| $PMG_{t\,1}$ | 0.108 [2.258] | 0.090 [1.813] | 0.168 [3.363] | 0.089 [1.349] | 0.111 [3.010] | 0.057 [1.347] |
| $PMG_{t\,2}$ | 0.190 [3.998] | 0.056 [1.130] | — | — | 0.171 [4.923] | 0.053 [1.292] |
| $PML_t$ | 0.050 [1.043] | 0.216 [4.268] | 0.232 [5.398] | 0.297 [5.232] | 0.139 [4.343] | 0.253 [6.727] |
| $PML_{t\,1}$ | 0.219 [4.449] | 0.086 [1.672] | 0.172 [3.711] | 0.029 [0.476] | 0.192 [5.661] | 0.032 [0.799] |
| $PML_{t\,2}$ | 0.148 [2.939] | 0.073 [1.394] | — | — | 0.105 [3.110] | 0.116 [2.912] |
| C | 0.012 [4.093] | 0.010 [3.208] | 0.015 [5.038] | 0.014 [3.776] | 0.012 [5.544] | 0.012 [4.698] |
| R-squared | 0.143 | 0.098 | 0.187 | 0.096 | 0.166 | 0.102 |
| Panel B. Quarterly Data Observations | | | | | | |
| | Over 1950.q1-1985.q4 | | Over 1986.q1-2015.q4 | | Over 1950.q1-2015.q4 | |
| | $PMG_{t+1}$ | $PML_{t+1}$ | $PMG_{t+1}$ | $PML_{t+1}$ | $PMG_{t+1}$ | $PML_{t+1}$ |
| $PMG_t$ | 0.395 [5.198] | -0.202 [-2.167] | 0.231 [2.663] | -0.066 [-0.595] | 0.319 [5.587] | -0.138 [-1.934] |
| $PML_t$ | 0.431 [6.222] | 0.117 [1.373] | 0.372 [5.068] | 0.253 [2.708] | 0.404 [8.048] | 0.185 [2.944] |
| C | 0.018 [2.602] | 0.018 [2.602] | 0.033 [3.917] | 0.038 [3.621] | 0.025 [4.609] | 0.045 [6.791] |
| R-squared | 0.272 | 0.059 | 0.187 | 0.076 | 0.227 | 0.062 |

Note. Vector Autoregression estimates. VAR lag order selection criteria is determined by SIC (Schwarz Information Criterion), t-statistics are reported in [ ]

This question is important since it relates to the efficiency of modeling stock returns as a unit. If there is no difference, then modeling stock returns as a unit is reasonable; otherwise, more elaborate models are required. We explored this question by performing predictable analysis. Our reasoning is that if there is no significant difference, then modeling stock returns as a unit should report forecasts no worse than modeling PMG and PML.

Merton's (1973) ICAPM suggests that the conditional expected return on the stock market should vary positively with the market's conditional variance. We used the ARCH-in-Mean (Engle, Lillien, and Robins, 1987) model as a benchmark to capture this risk-return tradeoff:

$$r^{\circ}_t = \delta_0 + \delta_1 r^{\circ}_{t-1} + \delta_2 h_t + e_t$$
$$h^2_t = \omega_0 + \omega_1 h^2_{t-1} + \omega_2 e^2_{t-1} + \omega_3 I(e_{t-1} < 0),$$

where $h^2_t$ is the conditional variance, $\delta_1 r^{\circ}_{t-1}$ is used to capture the possible autocorrelation in stock returns, and $\delta_2 h_t$ captures the time-varying risk premium; $\omega_3 I(e_{t-1}<0)$ is used to describe the possible "leverage effect" in volatility.

A vector autoregressive model of order $q$, VAR($q$) is used to describe the dynamics of PMG and PML:

$$PMG_t = C_g + \sum_{i=1}^{q} \alpha_{i,1} PMG_{t-i} + \sum_{i=1}^{q} \beta_{i,1} PML_{t-i} + \varepsilon^g_t$$
$$PML_t = C_l + \sum_{i=1}^{q} \alpha_{i,2} PMG_{t-i} + \sum_{i=1}^{q} \beta_{i,2} PML_{t-i} + \varepsilon^l_t.$$

The order $q$ was determined by the Schwarz information criterion (SIC). The forecasting values of stock returns were constructed from the following equation:

$$r^p_t = PMG^p_t - PML^p_t, \qquad (9)$$

where $PMG^p_t$ and $PML^p_t$ are forecasts reported by the VAR model.

### In-sample Evidence

The in-sample predictability of stock returns was performed over the whole sample. To be robust, in-sample fitting was also performed on two subsamples, 1950–1985 and 1986–2015.

Table 7 presents the estimations of the VAR(q) models together with the R-squares on PMG and PML. For monthly data observations, PML significantly contributed to forecasting PMG but not vice versa. For quarterly data observations, we found that both PML and PMG contributed to forecasting each other; however, PML contributed more than PMG. It is interesting to compare the R-squares in Table 3 with those in Table 7. The R-squares in Table 3 show that PML is more predictable than PMG when asymmetry is not taken into consideration. However, the R-squares in Table 8 show that PMG is more predictable than PML when asymmetry is considered. These findings further confirm that PML has a very significant impact on the dynamics of PMG.

Table 8 reports the in-sample R-squares on stock returns for both VAR(q) and ARCH-in-Mean. The R-squares reported by ARCH-in-Mean show very little predictability of stock returns. The ARCH-in-Mean model reports R-squares ranging from 0.16% to 0.27% for monthly returns and from 0.42% to 1.37% for quarterly returns. However, the R-squares reported by VAR(q) show high predictability for stock returns. The VAR(q) reports R-squares ranging from 1.17% to 1.89% for monthly returns and from 2.60% to 12.54% for quarterly returns. The sharp difference in the values of R-squares indicates that price extremes have additional information that cannot be explained by closing prices. These findings respond to the conjecture that modeling stock returns as a unit fails to discover the true dynamics of stock returns.

### Out-of-sample Evidence

A potential problem with in-sample predictability is overfitting. In a comprehensive study, Welch and Goyal (2008) showed that many macroeconomic variables, though delivering significant in-sample forecasts, performed poorly out of sample. They compared the mean squared error of the forecast $r^p_{t+1}$ with that of the sample mean return $r^m_{t+1}$ up to time t +1. Following Welch and Goyal (2008), the out-of-sample $R^2$, $R^2_{oos}$ is defined as

**Table 8** In-Sample R-squares Reported by VAR(q) and ARCH-in-Mean over Different Time Horizons

| $R_{in}^2$ | Panel A. Monthly Data Observations | | | Panel A. Quarterly Data Observations | | |
|---|---|---|---|---|---|---|
| | 1950.01-1985.12 | 1986.01-2015.12 | 1950.01-2015.12 | 1950.01-1985.12 | 1986.01-2015.12 | 1950.01-2015.12 |
| VAR(q) (%) | 1.89 | 1.17 | 1.57 | 12.54 | 2.60 | 6.76 |
| ARCH-in-Mean (%) | 0.32 | 0.27 | 0.16 | 1.37 | 0.42 | 0.82 |
| Predictability Ratio | 5.91 | 4.33 | 9.81 | 9.15 | 6.19 | 8.20 |

Note. This table reports the in-sample R-squares, $R_{in}^2$ reported by VAR(q) and ARCH-in-Mean. In-sample predictability is performed over different time horizons for the sake of being robust. Since ARCH tests report no heteroscedasticity in quarterly returns, thus we assume constant volatility for quarterly returns. Predictability ratio is calculated as the ratio of in-sample R-square of VAR(q) over the in-sample R-square of ARCH-in-Mean

Table 9 Out-of-Sample Forecasting Analysis

| Out-of-Sample | Panel A. Monthly Index Data | | | Panel B. Quarterly Index Data | | |
|---|---|---|---|---|---|---|
| | 1971.01-2015.12 | 1989.01-2015.12 | 1996.01-2015.12 | 1971Q1-2015Q4 | 1989Q1-2015Q4 | 1996Q1-2015Q4 |
| $R^2_{oos}$ (%) | 1.07* | 0.098 | 1.46 | 3.87** | 0.97* | 2.14* |
| CER (%) | 0.378 | 1.03 | 2.41 | 2.95 | 4.20 | 4.85 |
| $SR^p$ | 0.059 | 0.076 | 0.099 | 0.201 | 0.227 | 0.226 |
| $SR^{bh}$ | 0.037 | 0.085 | 0.068 | 0.060 | 0.138 | 0.109 |

Note. ***, **, * means respectively statistical significance at the level of 1%, 5% and 10%. The statistical significance of $R_{oos}^2$ is evaluated with the MSPE-adjusted statistic (Clark and West, 2007). The MSPE-adjusted statistic is conveniently calculated by first defining $f_{t+1} = (r_{t+1} - r^m_{t+1})^2 - [(r_{t+1} - r^p_{t+1})^2 - (r^m_{t+1} - r^p_{t+1})^2]$.
By regressing $f_{t+1}$ on a constant and calculating the t-statistic corresponding to the constant, a p-value for a one-sided (upper-tail) test is obtained with the standard normal distribution

$$R^2_{oos} = 1 - \sum\nolimits_{t=m+1}^{T}(r_t - r^p_t)^2 / \sum\nolimits_{t=m+1}^{T}(r_t - r^m_t)^2 \qquad (10)$$

Welch and Goyal (2008) found that $R^2_{oos}$ is generally less than zero for many return forecast variables.

We performed out-of-sample forecasting in a static way. Specifically, the total observations were divided into two subsamples. The first subsample, $\{x_t\}^M_{t=1}(x_t=(PMG_t, PML_t)$, was used to estimate the coefficients in the VAR model, and the remaining subsample, $\{x_t\}^M_{t=1}$, was used for out-of-sample forecasting evaluation:

$$r^p_{t+1} = PMG^p_{t+1} - PML^p_{t+1}, t = M, M+1, ..., T,$$

where $PMG^p_{t+1}$ and $PML^p_{t+1}$ are out-of-sample predictions reported by the VAR model.

Table 9 shows the out-of-sample R-squares, $R^2_{oos}$. For robustness, we performed out-of-sample forecasting on different subsamples. Three different time horizons were used: a long horizon from 1971 to 2015, a short horizon from 1996 to 2015, and midterm time horizon from 1989 to 2015. We found that all $R^2_{oos}$s were positive, indicating that the VAR(q) model consistently outperformed the simple historical mean in out-of-sample forecasting.

A limitation of the $R^2_{oos}$ measure is that it does not explicitly account for the risk borne by an investor over the out-of-sample period. To address this, following Campbell and Thompson (2008), we also calculated the realized utility gains for a mean-variance investor on a real-time basis. This exercise requires the investor to forecast the variance in stock returns. Similar to Campbell and Thompson (2008), we assumed that the investor estimates variance using a 10-year rolling window. A mean-variance investor who forecasts the equity premium using the historical average will decide at the end of period $t$ to allocate the following share of his or her portfolio to equities in period $t + 1$:

$$\omega_{0,t} = (1/\gamma)\left[(r^m_{t+1} - r^f_{t+1})/\sigma^2_{m,t+1}\right], \qquad (11)$$

where $r^f_{t+1}$ and $\sigma^2_{m,t+1}$ are, respectively, the risk-free rate and the rolling-window estimate of the variance in stock returns. Parameter $\gamma$ is the relative risk aversion.[2] Over the out-of-sample period, the investor realizes an average utility level of

$$v_0 = \mu_0 - 0.5\gamma\sigma^2_0, \qquad (12)$$

where $\mu_0$ and $\sigma^2_0$ are the sample mean and variance, respectively, over the out-of-sample period for the return on the benchmark portfolio formed using forecasts of the equity premium based on the historical sample mean. We then computed the average utility for the same investor when he or she forecasts the equity premium using our VAR regression model. He or she will choose an equity share of

$$\omega_{p,t} = (1/\gamma)\left[(r^p_{t+1} - r^f_{t+1})/\sigma^2_{m,t+1}\right] \quad (13)$$

and realize an average utility level of

$$v_p = \mu_p - 0.5\gamma\sigma^2_p, \quad (14)$$

where $\mu_p$ and $\sigma^2_p$ are the sample mean and variance, respectively, over the out-of-sample period for the return on the portfolio formed using forecasts of the equity premium based on our VAR model.

Table 10 Correlation Matrix. Monthly Observations

|      | PML   | PMG   | BM    | DE    | DFY   | DP    | DY    | EP    | INFL  | LTR   | LTY   | NTIS  | SVAR  | TBL   | TMS  |
|------|-------|-------|-------|-------|-------|-------|-------|-------|-------|-------|-------|-------|-------|-------|------|
| PML  | 1.00  |       |       |       |       |       |       |       |       |       |       |       |       |       |      |
| PMG  | -0.19 | 1.00  |       |       |       |       |       |       |       |       |       |       |       |       |      |
| BM   | 0.09  | 0.017 | 1.00  |       |       |       |       |       |       |       |       |       |       |       |      |
| DE   | 0.06  | 0.06  | 0.08  | 1.00  |       |       |       |       |       |       |       |       |       |       |      |
| DFY  | 0.19  | 0.25  | 0.28  | 0.22  | 1.00  |       |       |       |       |       |       |       |       |       |      |
| DP   | 0.01  | -0.04 | 0.88  | 0.32  | 0.15  | 1.00  |       |       |       |       |       |       |       |       |      |
| DY   | -0.07 | 0.03  | 0.87  | 0.32  | 0.15  | 0.99  | 1.00  |       |       |       |       |       |       |       |      |
| EP   | -0.02 | -0.09 | 0.81  | -0.37 | -0.00 | 0.76  | 0.76  | 1.00  |       |       |       |       |       |       |      |
| INFL | 0.14  | 0.04  | 0.42  | -0.17 | -0.10 | 0.28  | 0.27  | 0.39  | 1.00  |       |       |       |       |       |      |
| LTR  | -0.06 | 0.10  | -0.00 | -0.02 | 0.14  | -0.01 | -0.00 | 0.00  | -0.13 | 1.00  |       |       |       |       |      |
| LTY  | 0.13  | 0.11  | 0.42  | -0.08 | 0.51  | 0.26  | 0.26  | 0.30  | 0.41  | 0.04  | 1.00  |       |       |       |      |
| NTIS | -0.05 | -0.11 | 0.24  | 0.07  | -0.40 | 0.23  | 0.23  | 0.18  | 0.10  | -0.09 | -0.09 | 1.00  |       |       |      |
| SVAR | 0.54  | 0.08  | -0.10 | 0.15  | 0.32  | -0.08 | -0.11 | -0.18 | -0.14 | 0.14  | 0.01  | -0.25 | 1.00  |       |      |
| TBL  | 0.13  | 0.08  | 0.51  | -0.15 | 0.33  | 0.35  | 0.35  | 0.44  | 0.50  | 0.024 | 0.89  | -0.01 | -0.05 | 1.00  |      |
| TMS  | -0.04 | 0.05  | -0.31 | 0.16  | 0.27  | -0.26 | -0.25 | -0.36 | -0.29 | 0.00  | 0.01  | -0.16 | 0.14  | -0.44 | 1.00 |

We measured utility gain as the difference between Equation (14) and Equation (12) and multiplied this difference by 1200 (400) for monthly (quarterly) observations to express it as the average annualized percentage return. The utility gain (or certainty equivalent return, CER) can be interpreted as the portfolio management fee an investor would be willing to pay to have access to the additional information available in a predictive regression model relative to the information in the historical sample mean. We report the results for γ = 3. The results are qualitatively similar for other reasonable γ values.

The utility gains are also reported in Table 9. All utility gains are positive. For monthly returns, the utility gains range from 0.378% to 2.41%; for quarterly returns, the utility gains range from 2.95% to 4.85%. We also calculated the monthly Sharpe ratio of the portfolio, which is the mean portfolio return in excess of the risk-free rate divided by the standard deviation of the excess portfolio return. We used $SR^p$ and $SR^{bh}$ to represent, respectively, the Sharpe ratio of our constructed portfolio and the buy-and-hold portfolio. The Sharpe ratio showed the superior performance of our constructed portfolio over the simple buy-and-hold portfolio, except for the out-of-sample period January 1989 to December 2015 for monthly data observations.

### Potential Explanations

We demonstrated an asymmetry between PMG and PML in Section 3 and the economic value of the asymmetry in Section 4. This section explores whether this asymmetry can be explained by business cycle, technical indicators, skewness, or market sentiment.

### Business Cycle

A business cycle is an asymmetric economic condition that is long in expansion and short in recession. Thus, a potential explanation for asymmetry is that it is correlated with macroeconomic variables related to business cycle. Indeed, Chen, Roll and Ross (1986); Keim and Stambaugh (1986); Campbell and Shiller (1988); Fama and French (1988); Campbell (1991); Ferson and Harvey (1991); Lettau and Ludvigson (2001a, 2001b); and Li (2001) found evidence that the stock market can be predicted by variables related to business cycle, such as default spread, term spread, interest rate, inflation rate, dividend yield, consumption–wealth ratio, and surplus ratio.

For the monthly data, 13 representative business-cycle-related predictors were collected for the time period January 1950 to December 2015. The 13 economic variables are the following:

- *Book-to-market ratio, BM.* Ratio of book value to market value for the Dow Jones Industrial Average.
- *Dividend-payout ratio (log), D/E.* Difference between the log of dividends and the log of earnings.
- *Default yield spread, DFY.* Difference between BAA- and AAA-rated corporate bond yields.
- *Dividend-price ratio (log), D/P.* Difference between the log of dividends paid on the S&P 500 index and the log of stock prices (S&P 500 index), where dividends are measured using a one-year moving sum.
- *Dividend yield (log), D/Y.* Difference between the log of dividends and the log of lagged stock prices.
- *Earnings-price ratio (log), E/P.* Difference between the log of earnings on the S&P 500 index and the log of stock prices, where earnings are measured using a one-year moving sum.
- *Inflation, INFL.* Calculated from the CPI (all urban consumers).
- *Long-term return, LTR.* Return on long-term government bonds.
- *Long-term yield, LTY.* Long-term government bond yield.
- *Net equity expansion, NTIS.* Ratio of 12-month moving sums of net issues by NYSE-listed stocks to total end-of-year market capitalization of NYSE stocks.
- *Stock variance, SVAR.* Sum of squared daily returns on the S&P 500 index.
- *Treasury bill rate, TBL.* Interest rate on a three-month Treasury bill (secondary market).
- *Term spread, TMS.* Difference between the long-term yield and the Treasury bill rate.

For quarterly data, two more predictor variables were collected[3]:

- *Investment-to-capital ratio, IK.* Ratio of aggregate (private nonresidential fixed) investment to aggregate capital for the entire economy (Cochrane, 1991).
- *CAY.* CAY is defined as in Lettau and Ludvigson (2001a).

Tables 10 and 11 present, respectively, the summary statistics on correlations for monthly and quarterly data observations. Except *SVAR*, the results show low correlations between PML (PMG) and business-cycle-related variables. Tables 12 and Table 13 present, respectively, regression results for monthly and quarterly observations with business-cycle-related variables controlled. The regression results show that asymmetry cannot be explained by business-cycle-related variables.

**Technical Indicators**

Recent empirical literature has shown that some technical indicators are informative for forecasting stock returns. Huang et al. (2016) constructed two indicators from historical price—the mean reversion indicator and the good time indicator—and found that stock returns can be significantly predicted in both good and bad times. The mean reversion indicator and the good time indicator are defined as follows:

- *Mean reversion indicator, MRI.* This indicator has been found to be informative for predicting stock returns (Huang et al., 2016):

$$MRI_t = (r_{t-12 \to t} - u)/\sigma_{t-12 \to t},$$

where $r_{t-12 \to t}$ is the cumulative market return over the past year (from month t-11 to month t), u is the long-term mean (mean of the past 30 years), and $\sigma_{t-12 \to t}$ is the annualized moving standard deviation estimator (Mele, 2007).

- *Good time indicator, $I_{MA}$.* This indicator is the 200-day moving average. It takes a value of 1 when the S&P 500 index is above its 200-day moving average (Huang et al., 2016).

Table 14 reports the regression results for monthly data in panel A and for quarterly data in panel B. The results show that PML still significantly predicts PMG, even when MRI and IMA are controlled.

George and Hwang (2004) and Li and Yu (2012) showed, respectively, that 52-week high and historical high predict stock returns. Following George and Hwang (2004) and Li and Yu (2012), the 52-week high and historical high are presented as follows:

- *Nearness to the Dow 52-week high, $H^{52}$.* George and Hwang (2004) suggested that traders might use the 52-week high as an anchor when assessing the increment in stock value implied by new information. Suppose there are 250 trading days in 52 weeks; the nearness to the 52-week high was computed in this study as the ratio of the current S&P 500 index and its 250-day high:

$$H^{52}_t = p_t/p_{250,t},$$

where $p_t$ denotes the level of the S&P 500 index at the end of day t, and $p_{250,t}$ denotes the 250-day high at the end of day t.

Table 11 Correlation Matrix. Quarterly Observations

| | PML | PMG | BM | DE | DFY | DP | DY | EP | INFL | LTR | LTY | NTIS | SVAR | TBL | TMS | CAY | IK |
|---|---|---|---|---|---|---|---|---|---|---|---|---|---|---|---|---|---|
| PML | 1.00 | | | | | | | | | | | | | | | | |
| PMG | -0.03 | 1.00 | | | | | | | | | | | | | | | |
| BM | 0.14 | 0.01 | 1.00 | | | | | | | | | | | | | | |
| DE | 0.13 | 0.05 | 0.10 | 1.00 | | | | | | | | | | | | | |
| DFY | 0.25 | 0.23 | 0.33 | 0.23 | 1.00 | | | | | | | | | | | | |
| DP | 0.10 | -0.03 | 0.89 | 0.31 | 0.22 | 1.00 | | | | | | | | | | | |
| DY | -0.06 | 0.12 | 0.87 | 0.30 | 0.21 | 0.98 | 1.00 | | | | | | | | | | |
| EP | 0.00 | -0.07 | 0.78 | -0.43 | 0.04 | 0.73 | 0.72 | 1.00 | | | | | | | | | |
| INFL | 0.17 | 0.00 | 0.51 | -0.24 | 0.15 | 0.32 | 0.29 | 0.47 | 1.00 | | | | | | | | |
| LTR | 0.04 | 0.06 | -0.01 | -0.03 | 0.27 | 0.00 | 0.00 | 0.02 | -0.22 | 1.00 | | | | | | | |
| LTY | 0.10 | 0.09 | 0.47 | -0.08 | 0.49 | 0.37 | 0.36 | 0.41 | 0.56 | 0.04 | 1.00 | | | | | | |
| NTIS | -0.06 | 0.14 | 0.22 | 0.06 | -0.38 | 0.19 | 0.19 | 0.14 | 0.07 | -0.14 | -0.06 | 1.00 | | | | | |
| SVAR | 0.61 | -0.05 | -0.11 | 0.28 | 0.45 | -0.09 | -0.17 | -0.29 | -0.21 | 0.28 | -0.02 | -0.25 | 1.00 | | | | |
| TBL | 0.12 | 0.04 | 0.55 | -0.14 | 0.32 | 0.44 | 0.43 | 0.52 | 0.65 | -0.03 | 0.89 | 0.02 | -0.09 | 1.00 | | | |
| TMS | -0.07 | 0.09 | -0.29 | 0.15 | 0.24 | -0.25 | -0.23 | -0.35 | -0.33 | 0.14 | -0.01 | -0.14 | 0.15 | -0.47 | 1.00 | | |
| CAY | -0.09 | 0.08 | -0.14 | 0.21 | -0.03 | 0.12 | 0.14 | -0.04 | -0.17 | 0.14 | 0.30 | -0.06 | 0.03 | 0.17 | 0.22 | 1.00 | |
| IK | 0.15 | -0.04 | -0.01 | -0.33 | -0.12 | -0.18 | -0.20 | 0.07 | 0.35 | -0.02 | 0.32 | -0.02 | -0.02 | 0.51 | -0.49 | -0.16 | 1.00 |

**Table 12** Monthly Regression with Business-cycle Related Variables Controlled

| | $BM_t$ | $BM_{t+1}$ | $PML_t^F$ | $DE_t$ | $DE_{t+1}$ | $PML_t^F$ | $DFY_t$ | $DFY_{t+1}$ | $PML_t^F$ | $DP_t$ | $DP_{t+1}$ |
|---|---|---|---|---|---|---|---|---|---|---|---|
| $PML_t^F$ | | | | | | | | | | | |
| 0.136*** | | | | | | | | | | | |
| 0.133*** | 0.247* | | 0.137*** | 0.168 | | 0.135*** | 31.118 | | 0.135*** | 0.118 | |
| 0.127*** | 19.746*** | −19.627*** | 0.134*** | −0.262 | 0.436 | 0.130*** | 6.632 | 27.265 | 0.153*** | 17.205*** | −17.207*** |
| $PML_t^F$ | $DY_t$ | $DY_{t+1}$ | $PML_t^F$ | $EP_t$ | $EP_{t+1}$ | $PML_t^F$ | $INFL_t$ | $INFL_{t+1}$ | $PML_t^F$ | $LTR_t$ | $LTR_{t+1}$ |
| 0.139*** | 0.115 | | 0.136*** | 0.035 | | 0.138*** | −8.621 | | 0.153*** | 5.133*** | |
| 0.135*** | −0.007 | 0.123 | 0.100*** | 8.748*** | −8.811*** | 0.139*** | −9.618 | 1.608 | 0.144*** | 4.969*** | 2.983** |
| $PML_t^F$ | $LTY_t$ | $LTY_{t+1}$ | $PML_t^F$ | $NTIS_t$ | $NTIS_{t+1}$ | $PML_t^F$ | $SVAR_t$ | $SVAR_{t+1}$ | $PML_t^F$ | $TBL_t$ | $TBL_{t+1}$ |
| 0.134*** | 1.642 | | 0.136*** | −4.308** | | 0.112*** | 19.485** | | 0.133*** | 1.173 | |
| 0.123*** | 43.153*** | −41.711*** | 0.130* | 18.680* | −23.424** | 0.118*** | 25.238*** | −13.071 | 0.134*** | 37.059*** | −36.196*** |
| $PML_t^F$ | $TMS_t$ | $TMS_{t+1}$ | | | | | | | | | |
| 0.136*** | 0.722 | | | | | | | | | | |
| 0.142*** | −16.999* | 18.533** | | | | | | | | | |

Note. Our benchmark model is $PMG_{t+1}^F = C + PML_t^F + \varepsilon_{t+1}$, where $PMG^F$ and $PML^F$ are filtered observations. Filtered observations are used to alleviate the contamination of autocorrelations in PMG and PML. Regression with business-cycle related variables controlled is presented as follows,

$PMG_{t+1}^F = C + \alpha PML_t^F + \beta_1 M_t + \varepsilon_{t+1}$,

$PMG_{t+1}^F = C + PML_t^F + \beta_1 M_t + \beta_2 M_{t+1} + \varepsilon_{t+1}$,

where $M_t$ represents business-cycle related variable. The constant C is not reported in the table for space-saving. ***, **, * mean respectively significance at the level of 1%, 5% and 10%

**Table 13** Quarterly Regression with Business-cycle Related Variables Controlled

| $PML^F_t$ | $BM_t$ | $BM_{t+1}$ | $PML^F_t$ | $DE_t$ | $DE_{t+1}$ | $PML^F_t$ | $DFY_t$ | $DFY_{t+1}$ | $PML^F_t$ | $DP_t$ | $DP_{t+1}$ |
|---|---|---|---|---|---|---|---|---|---|---|---|
| 0.292*** | | | | | | | | | | | |
| 0.285*** | 0.030 | | 0.290*** | 0.017 | | 0.277*** | 2.815** | | 0.287*** | 0.021 | |
| 0..234*** | 1.158*** | -1.150*** | 0.310*** | 0.095** | -0.087** | 0.310*** | 6.973*** | -4.833* | 0.248*** | 0.851*** | 0.851*** |
| $PML^F_t$ | $DY_t$ | $D_{t+1}$ | $PML^F_t$ | $EP_t$ | $EP_{t+1}$ | $PML^F_t$ | $INFL_t$ | $INFL_{t+1}$ | $PML^F_t$ | $LTR_t$ | $LTR_{t+1}$ |
| 0.301*** | 0.023* | | 0.291*** | 0.011 | | 0.292*** | 0.000 | | 0.294*** | 0.212** | |
| 0.440*** | 0.252** | -0.231** | 0.239*** | 0.195*** | -0.196*** | 0.294*** | 0.304 | -0.513 | 0.295*** | 0.212** | -0.018 |
| $PML^F_t$ | $LTY_t$ | $LTY_{t+1}$ | $PML^F_t$ | $NTIS_t$ | $NTIS_{t+1}$ | $PML^F_t$ | $SVAR_t$ | $SVAR_{t+1}$ | $PML^F_t$ | $TBL_t$ | $TBL_{t+1}$ |
| 0.288*** | 0.123 | | 0.290*** | -0.440 | | 0.261*** | 0.784 | | 0.291*** | 0.015 | |
| 0.284*** | 0.665 | -0.550 | 0.286*** | -0.128 | -0.335 | 0.280*** | 1.720*** | -2.435*** | 0.289*** | 0.257 | -0.252 |
| $PML^F_t$ | $TMS_t$ | $TMS_{t+1}$ | $PML^F_t$ | $CAY_t$ | $CAY_{t+1}$ | $PML^F_t$ | $IK_t$ | $IK_{t+1}$ | | | |
| 0.297*** | 0.396 | | 0.305*** | 0.691*** | | 0.302*** | -1.544 | | | | |
| 0.296*** | 0.126 | 0.319 | 0.315*** | 4.026*** | -3.414*** | 0.295*** | 5.067 | -6.779 | | | |

Note. Our benchmark model is $PMG^F_{t+1}=C+PML^F_t+\varepsilon_{t+1}$, where $PMG^F$ and $PML^F$ are filtered observations. Filtered observations are used to alleviate the contamination of autocorrelations in PMG and PML. Regression with business-cycle related variables controlled is presented as follows,

$PMG^F_{t+1} = C + \alpha PML^F_t + \beta_1 M_t + \varepsilon_{t+1}$,
$PMG^F_{t+1} = C + PML^F_t + \beta_1 M_t + \beta_2 M_{t+1} + \varepsilon_{t+1}$,

where $M_t$ represents business-cycle related variable. The constant C is not reported in the table for space-saving. ***, **, * mean respectively significance at the level of 1%, 5% and 10%

- *Nearness to the historical high, $H^{max}$*. Following Li and Yu (2012), nearness to the historical high was calculated as the ratio of the current S&P 500 index and its historical high:

$$H^{max}_t = p_t/p_{max,t},$$

where $p_{max,t}$ denotes its historical high at the end of day t.

Table 15 reports the regression results, which are mixed. For monthly data observations in panel A, the results show that asymmetry can be explained by 52-week-high

**Table 14** Regression with Mean Reversion Indicator and Good Time Indicator Controlled

| Panel A. Monthly Data Observations | | | | |
|---|---|---|---|---|
| $PML^F_t$ | $I_{MA,t}MRI_t$ | $(1-I_{MA,t})MRI_t$ | $I_{MA,t+1}MRI_{t+1}$ | $(1-I_{MA,t+1})MRI_{t+1}$ |
| 0.136*** | | | | |
| 0.105*** | -0.178*** | -0.174** | | |
| 0.130*** | -1.071*** | -1.147*** | 0.981*** | 1.063*** |
| Panel B. Quarterly Data Observations | | | | |
| $PML^F_t$ | $I_{MA,t}MRI_t$ | $(1-I_{MA,t})MRI_t$ | $I_{MA,t+1}MRI_{t+1}$ | $(1-I_{MA,t+1})MRI_{t+1}$ |
| 0.292*** | | | | |
| 0.259*** | -0.007 | -0.008 | | |
| 0.396*** | -0.059*** | -0.073*** | 0.077*** | 0.106*** |

Note. Our benchmark model is $PMG^F_{t+1}=C+PML^F_t+\varepsilon_{t+1}$, where $PMG^F$ and $PML^F$ are filtered observations. Filtered observations are used to alleviate the contamination of autocorrelations in PMG and PML. Regression with mean reversion indicator controlled is presented as follows,

$PMG^F_{t+1} = C + \alpha PML^F_t + \beta_1 I_{MA,t}MRI_t + \beta_2(1-I_{MA,t}MRI_t) + \varepsilon_{t+1}$,
$PMG^F_{t+1} = C + \alpha PML^F_t + \beta_1 I_{MA,t}MRI_t + \beta_2(1-I_{MA,t}MRI_t) + \beta_3 I_{MA,t+1}MRI_{t+1} + \beta_4(1-I_{MA,t+1}MRI_{t+1}) + \varepsilon_{t+1}$.,
This regression follows Huang et al. (2015) who use the following state-dependent regression to predict stock returns, $r_{t+1}$
$r_{t+1} = C + \beta_1 I_{MA,t}MRI_t + \beta_2(1-I_{MA,t}MRI_t) + \varepsilon_{t+1}$
The constant C is not reported in the table for space-saving. ***, **, * mean respectively significance at the level of 1%, 5% and 10%

**Table 15** Regression with 52-week High and Historical High Controlled

| Panel A. Monthly Data Observations | | | | |
|---|---|---|---|---|
| $PML^F_t$ | $H^{52}_t$ | $H^{max}_t$ | $H^{52}_{t+1}$ | $H^{max}_{t+1}$ |
| 0.136*** | | | | |
| 0.053 | -2.859*** | -0.464** | | |
| 0.021 | 1.865*** | -25.152*** | -5.607** | 25.926*** |
| Panel B. Quarterly Data Observations | | | | |
| $PML^F_t$ | $H^{52}_t$ | $H^{max}_t$ | $H^{52}_{t+1}$ | $H^{max}_{t+1}$ |
| 0.292*** | | | | |
| 0.177** | -0.178 | -0.059 | | |
| 0.092* | -0.032 | -1.247*** | -0.345* | 1.387*** |

Note. Our benchmark model is $PMG^F_{t+1}=C+PML^F_t+\varepsilon_{t+1}$, where $PMG^F$ and $PML^F$ are filtered observations. Filtered observations are used to alleviate the contamination of autocorrelations in PMG and PML. Regression with mean reversion indicator controlled is presented as follows,
$PMG^F_{t+1} = C + \alpha PML^F_t + \beta_1 H^{52}_t + \beta_2 H^{max}_t + \varepsilon_{t+1}$,
$PMG^F_{t+1} = C + \alpha PML^F_t + \beta_1 H^{52}_t + \beta_2 H^{max}_t + \beta_3 H^{52}_{t+1} + \beta_4 H^{max}_{t+1} + \varepsilon_{t+1}$.,
The constant C is not reported in the table for space-saving. ***, **, * mean respectively significance at the level of 1%, 5% and 10%

and historical-high indicators. The coefficients of filtered bad extreme returns decreased from 0.136 to 0.053 and to 0.021, and became insignificant. For quarterly data observations in panel B, the results show that the 52-week high and historical high only partly explained the asymmetry in covariances.

A superior explanation of 52-week high and historical high in relation to asymmetry pertains to the similarities among PML, $H^{52}_t$, and $H^{max}_t$. All of these three indicators are constructed from high price. Actually, the correlation between PML and $H^{52}_t$ ($H^{max}_t$) was -0.606 (-0.449) for monthly data and -0.817 (-0.594) for quarterly data. However, the out-of-sample R-squares in Li and Yu (2012) were reported to be 0.1% for monthly data and 0.8% for quarterly data when both nearness to historical high and nearness to 52-week high were used as predictors, which are much smaller than our results.

#### Market Sentiment

Baker and Wurgler (2006) constructed an investor sentiment index and found that it had strong forecasting power for a large number of cross-sectional stock returns. Stambaugh, Yu, and Yuan (2012) found that investor sentiment predicted the short legs of long–short investment strategies. Baker, Wurgler, and Yuan (2012) provided further international evidence for the cross-section forecasting power of investor sentiment. Huang et al. (2015) found that an aligned sentiment index had much greater power in predicting the aggregate stock market than the Baker and Wurgler (2006) index.

We collected both the sentiment index of Baker and Wurgler (2006) and the aligned sentiment index of Huang et al. (2015) and investigated whether asymmetry could be explained by sentiment.[4] The results are presented in Table 16. Consistently, we found that asymmetry could not be explained by sentiment.

#### Skewness

Recent empirical results have shown that skewness predicts stock returns. Among others, Boyer, Mitton, and Vorkink (2010) found that expected idiosyncratic skewness

Table 16 Regression with Sentiment Index Controlled

| $PML_t^F$ | $SI_t$ | $SI_{t+1}$ | $PML_t^F$ | $ASI_t$ | $ASI_{t+1}$ |
|---|---|---|---|---|---|
| 0.136*** | | | | | |
| 0.162*** | -0.018 | | 0.163*** | -0.012 | |
| 0.151*** | 1.225*** | -1.256*** | 0.164** | 0.661*** | -0.688*** |

Note;The sentiment index is available from the homepage of Guofu Zhou: http://apps.olin.wustl.edu/faculty/zhou/. Only monthly sentiment index data is only available for the sample period from July 1965 to December 2014. Our benchmark model is $PMG^F_{t+1}=C+\alpha PML^F_t+\varepsilon_{t+1}$, where $PMG^F_{t+1}$ and $PML^F_t$ are filtered observations. Regression with sentiment index controlled is presented as follows,
$PMG^F_{t+1} = C + \alpha PML^F_t + \beta_1 IS_t + \beta_2 IS_{t+1} + \varepsilon_{t+1}$;
where $IS_t=SI_t$ or $ASI_t$. $SI_t$ and $ASI_t$ represent respectively the sentiment index of Baker and Wurgler (2006) and the aligned sentiment index of Huang et al. (2015). The constant C is not reported in the table for space-saving. ***, **, * mean respectively significance at the level of 1%, 5% and 10% and returns were negatively correlated. Amaya and Vasquez (2015) found that skewness from high-frequency data predicted the cross-section of stock returns. Using data on individual stock options, Rehman and Vilkov (2012) found that the currently observed option implied that ex ante skewness is positively related to future stock returns. We constructed the skewness indicator as follows:

$$SK_t = \sum_{i=1}^{200}[(r_{t+1-i}-u_t)/\sigma_t]^3/200$$
$$u_t = \sum_{i=1}^{200}r_{t+1-i}/200$$
$$\sigma_t = \sum_{i=1}^{200}(r_{t+1-i}-u_t)^2/200,$$

where $SK_t$ is the skewness indicator, and $r_t$s are daily log returns. Similar to Huang et al. (2016), we used 200 as the moving window length. Table 17 reports the results, which show that asymmetry cannot be explained by skewness.

## Global Evidence

To see if the asymmetry discovered in this study is only applicable to the US S&P 500 stock index, we also performed a comprehensive empirical study on the main global stock indices. For Asian countries, this included the Chinese Shanghai Stock Exchange Composite index (SSEC), Hongkong Hangseng Index (HS), Taiwan Stock Exchange Corporation index (TSEC), Singapore Strait Times Index (ST), Japanese Nikkei 225 Stock Average Index (NIKKEI 225, NK), Korea Stock Exchange Kospi Index (KOSPI). For European countries, we used the British Financial Times Stock Exchange 100 Index (FTSE 100, FT), German Deutscher Aktien Index (DAX), and French Cotation Assistee en Continu 40 Index (CAC 40, CAC). It also included the National Association of

Table 17 Regression with Skewness Controlled

| Panel A: Monthly Data Observations | | | Panel B: Quarterly Data Observations | | |
|---|---|---|---|---|---|
| $PML_t^F$ | $SK_t$ | $SK_{t+1}$ | $PML_t^F$ | $SK_t$ | $SK_{t+1}$ |
| 0.136*** | | | 0.292*** | | |
| 0.146*** | 0.055 | | 0.300*** | 0.013 | |
| 0.144*** | -0.102 | 0.176 | 0.298*** | 0.010 | 0.004 |

Note;$SK_t$ is the skewness indicator. Our benchmark model is $PMG^F_{t+1}=C+\alpha PML^F_t+\varepsilon_{t+1}$, where $PMG^F_{t+1}$ and $PML^F_t$ are filtered observations. Regression with skewness controlled is presented as follows,
$PMG^F_{t+1} = C + \alpha PML^F_t + \beta_1 SK_t + \beta_2 SK_{t+1} + \varepsilon_{t+1}$
The constant C is not reported in the table for space-saving. ***, **, * mean respectively significance at the level of 1%, 5% and 10%

Securities Dealers Automated Quotations Index (NASDAQ). All data sets were downloaded from www.finance.yahoo.com.

Table 18 reports the Granger causality testing results on PMG and PML when stock returns were decomposed with high price extremes. Almost unanimously, we found that PML Granger-caused PMG but not vice versa. This global evidence indicates that the asymmetric reactions between PMG and PML are general. For robustness, Granger causality tests were also performed when stock returns were decomposed with low price extremes; the results were similar. We didn't report the results for saving space.

## Conclusions

It is well known that price extremes are valuable for estimating and forecasting the volatility of financial assets. However, little is known about whether price extremes contribute to forecasting asset returns.

This study decomposed asset returns with price extremes into potential maximum gains (PMG) and potential maximum losses (PML) and empirically investigated the relationship between the two. We found significant asymmetry between PMG and PML. PML had a large impact on the time series dynamics of PMG but not vice versa. This asymmetry cannot be explained by macroeconomic variables, technical indicators, market sentiment, or skewness. We also explored the economic value of this asymmetry and found that investors can significantly improve their utility gains if this asymmetry

**Table 18** Granger Causality Tests on PMG and PML: Decomposition with High Price Extremes

|  | Panel A: Monthly Data Observations | | | Panel B: Quarterly Data Observations | | |
|---|---|---|---|---|---|---|
| Lags | 2 | 4 | 6 | 2 | 4 | 6 |
| HS: PMG /→ PML | 0.069 | 0.033 | 0.068 | 0.078 | 0.551 | 0.797 |
| HS: PML /→ PMG | 0.000 | 0.000 | 0.001 | 0.175 | 0.132 | 0.155 |
| ST: PMG /→ PML | 0.924 | 0.524 | 0.417 | 0.726 | 0.730 | 0.811 |
| ST: PML ST /→ PMG | 0.000 | 0.000 | 0.000 | 0.000 | 0.000 | 0.000 |
| NK: PMG /→ PML | 0.463 | 0.842 | 0.904 | 0.815 | 0.601 | 0.453 |
| NK: PML /→ PMG | 0.000 | 0.000 | 0.000 | 0.024 | 0.075 | 0.210 |
| FT: PMG /→ PML | 0.894 | 0.919 | 0.304 | 0.478 | 0.364 | 0.501 |
| FT: PML /→ PMG | 0.000 | 0.000 | 0.000 | 0.006 | 0.031 | 0.028 |
| NASDAQ:PMG /→ PML | 0.001 | 0.004 | 0.055 | 0.614 | 0.022 | 0.124 |
| NASDAQ:PML /→ PMG | 0.000 | 0.000 | 0.000 | 0.000 | 0.000 | 0.000 |
| SSEC: PMG /→ PML | 0.056 | 0.016 | 0.006 | | | |
| SSEC: PML /→ PMG | 0.032 | 0.080 | 0.310 | | | |
| TSEC: PMG /→ PML | 0.719 | 0.655 | 0.218 | | | |
| TSEC: PML /→ PMG | 0.000 | 0.000 | 0.000 | | | |
| KOSPI: PMG /→ PML | 0.055 | 0.334 | 0.059 | | | |
| KOSPI: PML /→ PMG | 0.000 | 0.000 | 0.000 | | | |
| DAX : PMG /→ PML | 0.508 | 0.944 | 0.690 | | | |
| DAX: PML /→ PMG | 0.000 | 0.000 | 0.000 | | | |
| CAC: PMG /→ PML | 0.271 | 0.328 | 0.269 | | | |
| CAC: PML /→ PMG | 0.000 | 0.000 | 0.000 | | | |

Note: $X /\rightarrow Y$ means the null hypothesis that $X$ does not Granger-causes $Y$. This table reports the $p$-values of the $F$-statistics. To make sure that there are enough data observations to perform Granger causality tests, therefore, for quarterly data observations, we only perform Granger causality tests on SP500, NASDAQ, FTSE100, HS, NK and ST.

is used in investment decisions. Moreover, this asymmetry was found to be quite general across the main global stock markets.

This study's findings have some interesting implications. First, there are elaborate intrinsic structures in the dynamics of asset returns, and these structures can hardly be captured by the univariate time series modeling technique. Thus, more subtle models are needed to describe the time series dynamics of asset returns. Second, the information contained in price extremes is valuable for asset pricing. Our future efforts will focus on incorporating price extremes into asset returns modeling and asset pricing.

## Endnotes

[1]The website only provides index data beginning in January 1950.

[2]Following Campbell and Thompson (2008), we constrained the portfolio weight on stocks to lie between 0% and 150% (inclusive) each month, so that $\omega_{0,t}=0$ ($\omega_{0,t}=1.5$) if $\omega_{0,t}<0$ ($\omega_{0,t}>1.5$) in Equation (11).

[3]All of these 15 economic variables are available on Amit Goyal's website: http//www.hec.unil.ch/agoyal/.

[4]The sentiment index is available on Guofu Zhou's website: http.apps.olin.wustl.edu/faculty/zhou/. Only monthly sentiment index data are available for the sample period from July 1965 to December 2014.

### Abbreviations
CARR: Conditional AutoRegressive Range; CER: Certainty Equivalent Return; GARCH: Generalized AutoRegressive Conditional Heteroskedasticity; ICAPM: Intertemporal Capital Asset Pricing Model; OVR: Overnight Return; PMG: Potential Maximum Gain; PML: Potential Maximum Loss; VAR: Vector Autoregressive Model

### Acknowledgements
We thank the anonymous referees. Their comments and suggestions have greatly improved our paper.

### Funding
This research is supported by National Natural Science Foundation of China under Grant No.71401033, and Program for Young Excellent Talents, UIBE under Grant No. 15YQ08.

### Availability of data and materials
All the data observations used in this paper are downloaded from the finance subdirectory of the website "Yahoo.com"

### Authors' contributions
SY contribution: He is the corresponding author and provides the most of the views and ideas of this paper. XH contribution: He is the first author and assists the corresponding author to complete the construction and writing of this paper. Both authors read and approved the final manuscript.

### Competing interests
The authors declare that we have no competing interests.

### Publisher's Note
Springer Nature remains neutral with regard to jurisdictional claims in published maps and institutional affiliations.

### Author details
[1]School of Banking and Finance, University of International Business and Economics, Beijing 100029, China. [2]Academy of Mathematics and Systems Science, Chinese Academy of Sciences, Beijing 100190, China.

### References
Amaya D, Vasquez A (2015) Skewness from High-Frequency Data Predicts the Cross-Section of Stock Returns[J]. J Financ Econ 118:135–167
Andersen T, Bollerslev T, Diebold F, Vega C (2003) Micro effects of macro announcements. real time price discovery in foreign exchange[J]. Am Econ Rev 93(1):38–62
Baker M, Wurgler J (2006) Investor Sentiment and the Cross-Section of Stock Returns[J]. J Financ 61(4):1645–1680
Baker M, Wurgler J, Yuan Y (2012) Global, local, and contagious investor sentiment?[J]. J Financ Econ 104(37):272–287

Beckers S (1983) Variance of Security Price Returns Based on High, Low and Closing Prices[J]. J Bus 56(1):97–112
Boyer B, Mitton T, Vorkink K (2010) Expected Idiosyncratic Skewness[J]. Rev Financ Stud 23(1):169–202
Brandt M, Jones C (2006) Volatility Forecasting with Range-Based EGARCH Models[J]. J Bus Econ Stat 24(4):470–486
Campbell JY, Shiller RJ (1988) Stock Prices, Earnings And Expected Dividends[J]. Journal of Finance 43(3):661–676
Campbell JY, Thompson SB (2008) Predicting Excess Stock Returns Out of Sample. Can Anything Beat the Historical Average?[J]. Rev Financ Stud 21(4):1509–1531
Campbell JY (1991) A variance decomposition model for stock returns[J]. Econ J 101(405):157–179
Chen NF, Roll R, Ross SA (1986) Economic forces and the stock market[J]. J Bus 59(3):383–403
Chou RY (2005) Forecasting Financial Volatilities with Extreme Values. The Conditional Autoregressive Range (CARR) Model[J]. J Money Credit Bank 37(3):561–582
Clark TE, West KD (2007) Approximately normal tests for equal predictive accuracy in nested models[J]. Nber Tech Working Pap 138(1):291–311
Cochrane JH (1991) Production-Based Asset Pricing and the Link between Stock Returns and Economic Fluctuations[J]. J Financ 46(1):209–237
Dangl T, Halling M (2012) Predictive regressions with time-varying coefficients[J]. J Financ Econ 106(1):157–181
Engle RF, Lilien DM, Robins RP (1987) Estimating Time Varying Risk Premia in the Term Structure. The Arch-M Model.[J]. Econometrica 55(2):391–407
Fama EF, French KR (1988) Dividend yields and expected stock returns ☆[J]. J Financ Econ 22(1):3–25
Ferson WE, Harvey CR (1991) The Variation of Economic Risk Premiums[J]. J Pol Econ 99(2):385–415
Garman MB, Klass MJ (1980) On the Estimation of Price Volatility from Historical Data[J]. J Bus 53(1):67–78
George TJ, Hwang C (2004) The 52-Week High and Momentum Investing[J]. J Financ 59(5):2145–2176
Granger CWJ (1969) Investigating Causal Relations by Econometric Models and Cross-spectral Methods. Econometrica[J]. Econometrica 37(3):424–438
Henkel SJ, Martin JS, Nardari F (2011) Time-Varying Short-Horizon Return Predictability[J]. J Financ Econ 99(3):560–580
Hong H, Lim T, Stein JC (2000) Bad News Travels Slowly. Size, Analyst Coverage, and the Profitability of Momentum Strategies[J]. J Financ 55(1):265–295
Hong H, Stein JCA (1999) Unified Theory of Underreaction, Momentum Trading, and Overreaction in Asset Markets[J]. J Financ 54(6):2143–2184
Huang D, Jiang F, Tu J, Zhou G (2015) Investor Sentiment Aligned. A powerful predictor of stock returns[J]. Rev Financ Stud 28(3):791–837
Huang, Dashan and Jiang, Fuwei and Tu, Jun and Zhou, Guofu, Forecasting Stock Returns in Good and Bad Times. The Role of Market States (January 31, 2016). 27th Australasian Finance and Banking Conference 2014 Paper; Asian Finance Association (AsianFA) 2016 Conference. Available at SSRN. http://ssrn.com/abstract=2188989 or http://dx.doi.org/10.2139/ssrn.2188989
Kahneman D, Tversky A (1979) Prospect theory. An analysis of decision under risk.[J]. Econometrica 47(2):263–291
Keim DB, Stambaugh RF (1986) Predicting Returns in Stock and Bond Markets[J]. J Financ Econ 17(2):357–390
Kunitomo N (1992) Improving the Parkinson Method of Estimating Security Price Volatilities[J]. J Bus 65(2):295–302
Lettau M, Ludvigson S (2001a) Consumption, Aggregate Wealth, and Expected Stock Returns[J]. J Financ 56(3):815–849
Lettau M, Ludvigson S (2001b) Resurrecting the (C)CAPM. a cross-sectional test when risk premia are time-varying[J]. J Pol Econ 109(109):1238–1287
Li J, Yu J (2012) Investor attention, psychological anchors, and stock return predictability[J]. J Financ Econ 104(2):401–419
Li Y (2001) Expected Returns and Habit Persistence[J]. Rev Financ Stud 14(3):861–899
Martens M, Dijk van D (2007) Measuring volatility with the realized range[J]. J Econometrics 138(1):181–207
Mele A (2007) Asymmetric stock market volatility and the cyclical behavior of expected returns[J]. J Financ Econ 86(2):446–478
Merton RC (1973) An Intertemporal Capital Asset Pricing Model.[J]. Econometrica 41(5):867–887
Nguyen VH, Claus E (2013) Good news, bad news, consumer sentiment and consumption behavior[J]. J Econ Psychol 39(39):426–438
Parkinson M (1980) The extreme value method for estimating the variance of the rate of return[J]. J Bus 53(1):61–65
Rapach DE, Strauss JK, Zhou G (2010) Out-of-Sample Equity Premium Prediction. Combination Forecasts and Links to the Real Economy[J]. Rev Financ Stud 23(2):821–862
Rehman Z, Vilkov G. Risk-Neutral Skewness. Return Predictability and Its Sources (March 13, 2012). Available at SSRN. http://ssrn.com/abstract=1301648 or http://dx.doi.org/10.2139/ssrn.1301648
Rogers LCG, Satchell SE (1991) Estimating Variance From High, Low and Closing Prices[J]. Annals Appl Probability 1(4):504–512
Alizadeh S, Brandt MW, Diebold FX (2002) Rang-based estimation of stochastic volatility models[J]. J Financ 57(3):1047–1091
Stambaugh RF, Yu J, Yuan Y (2012) The short of it. Investor sentiment and anomalies[J]. J Financ Econ 104(2):288–302
Veronesi P (1999) Stock Market Overreaction to Bad News in Good Times. A Rational Expectations Equilibrium Model[J]. Rev Financ Stud 12(5):975–1007
Welch I, Goyal A (2008) A Comprehensive Look at The Empirical Performance of Equity Premium Prediction[J]. Rev Financ Stud 21(4):1455–1508(54)
Wiggins JB (1991) Empirical tests of the bias and efficiency of the extreme-value variance estimator for common stocks[J]. J Bus 12(1):417–432
Yang D, Zhang Q (2000) Drift independent volatility estimation based on high, low, open and close prices[J]. J Bus 73(3):477–491